\documentstyle[psfig]{l-school}

\begin{document}
\markboth{Patrick Petitjean}{QSO Absorption Line Systems}
%
\setcounter{part}{15}
%
\title{QSO Absorption Line Systems}
%
\author{Patrick Petitjean}
%
\institute{Institut d'Astrophysique de Paris\\
98bis Boulevard Arago, 75014 Paris \\
France\\
petitjean@iap.fr
\\
}
\maketitle
%
%
%
%
%
It is difficult to describe in a few pages the numerous specific techniques
used to study absorption lines seen  in QSO spectra 
and to review even rapidly the field of research based on their
observation and analysis. What follows is therefore a pale introduction
to the invaluable contribution of these studies to our knowledge of the 
gaseous component of the Universe and its cosmological evolution.
A rich bibliography is given which, although not complete, will be hopefully 
useful for further investigations. Emphasis will be laid on the impact of
this field on the question of the formation and evolution of galaxies.
\section{INTRODUCTION}
\begin{figure}
\flushleft{\vskip -2.cm
\vbox{
\hskip -3.2cm
\psfig{figure=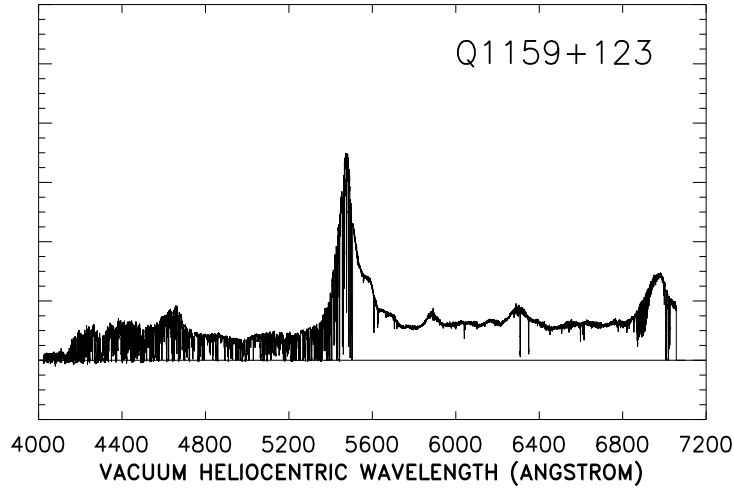,height=11.cm,width=15.cm,angle=0}
}}  
\vskip -3cm
\caption{QSO spectrum from Songaila \cite{son98} obtained
with the high-resolution spectrograph on the Keck telescope. The emission
redshift of the quasar Q1159+123 is $z_{\rm em}$~=~3.502; its
strongest emission lines are Ly$\beta$1025/O{\sc vi}1036 
at 4615~\AA, Ly$\alpha$1215 at 5473~\AA, N{\sc v}1240 at 5582~\AA, 
Si{\sc iv}1400 at 6300~\AA~ and C{\sc iv}1549 at 6974~\AA.
To the blue of the quasar Ly$\alpha$ emission line, 
the number of absorption lines is very large. Most of the lines are
attributed to Ly$\alpha$1215 from neutral hydrogen in intergalactic clouds;
when the quasar is at large redshift like this one, the corresponding
Ly$\beta$ lines are redshifted in the optical band and observed, confirming
the identification. The break in the continuum at 4150~\AA~
reveals the presence of a cloud optically thick at the Lyman limit 
(912~\AA~ at rest). To the red of the quasar Ly$\alpha$ emission line,
the number of absorption lines is smaller.
These lines can be attributed to different ionization stages of metals.
Since most of the strongest absorption lines are in the near UV at rest,
a wide range of ionization stages is accessible for the analysis of
the physical conditions in the gas.
}
\label{Fig1}
\end{figure}
Although steady progress is made towards detecting faint objects
in emission, the high-redshift objects detected this way up to now are 
drawn from a particular population of powerful emitters (e.g. \cite{ste98}).
On the contrary, absorption may reveal
standard objects such as a normal galaxy or intergalactic gaseous clouds
close enough to the line of sight to QSOs all over the range
of accessible redshifts from zero to the highest QSO redshift
$z$~=~4.897 \cite{schneider91}. A typical spectrum is shown in
Fig.~\ref{Fig1}.
Absorption line systems observed in QSO spectra are generally
classified into three categories: \par\noindent
(1) The metal-line systems in which 
a large number of elements, in different ionization stages, is observed,
from C~{\sc i} to C~{\sc iv} or O~{\sc i} to O~{\sc vi}. 
The Lyman limit systems (LLS) are optically thick at 
the H~{\sc i} Lyman limit (912~\AA) and have therefore
H~{\sc i} column densities $N$(H~{\sc i})~$>$~2$\times$10$^{17}$~cm$^{-2}$.
They are closely related, at any redshift, to galaxies. The latter have been 
detected by direct imaging and follow-up spectroscopy
at low and intermediate redshift\cite{ber91,gui97,ste93}. 
Studying their number density and physical
properties (kinematics, ionization state, abundances)
is then a unique tool to trace galaxy formation.
Among these systems, the rare damped systems, characterized by a very
large H~{\sc i} column density
($N$(H~{\sc i})~$>$~2$\times$10$^{20}$~cm$^{-2}$) are often
considered to be associated with galactic disks or protogalactic 
disks\cite{wol88}.\par\noindent
(2) The Ly$\alpha$ lines with no metal lines detected at the same redshift,
are of intergalactic origin at high
redshift (e.g. \cite{sar80}) but could somehow be associated
with galaxies at low redshift\cite{lan98}.
Information about these systems (column density, Doppler parameter,
kinematics, clustering) is derived from line-profile fitting. 
The latter requires high-resolution high-quality spectra that are
difficult to obtain on 4m-class telescopes. Recent studies of 
absorptions coincident in redshift in the spectra of
gravitationally lensed QSO images \cite{sme95} or QSO pairs
(e.g. \cite{dod98}) have shown that the mean radius of the absorbers is
surprizingly large (up to 500~kpc). The idea that the Ly$\alpha$ gas could
trace the potential wells of dark matter \cite{cen94} and 
reveal its characteristical filamentary structure has been 
successfully investigated using $N$-body simulations \cite{muc96,pet95}.
\par\noindent
(3) The broad absorption line (BAL) systems are characterized by impressive 
absorption troughs from different ions of low and high excitation,
extending from 0 up to 60000~km~s$^{-1}$ outflow velocity 
relative to the QSO emission redshift. It is widely accepted that the gas
is  very close to the centre of the QSO host-galaxy and 
may be part of the broad emission line region (\cite{wam95, turn96}, see
also \cite{ham97}). They
thus are intimately related to the AGN phenomenon. Although
AGN activity is an important factor of galaxy formation, the role of
BAL systems in galaxy formation is unclear at the moment.
This is the bad reason why they are not described in this lecture.\par
\section{ABSORPTION LINES}
In this Section, we summarize a few basic and classical results on
the formation and characteristics of absorption lines, that are important
to keep in mind when discussing the scientific 
issues presented in the following Sections.\par\noindent
\subsection{Definitions}
An absorption line has an equivalent width $w_{\rm obs}$
in \AA~ defined as:
\begin{equation}
w_{\rm obs} = \int {I_{\rm c}-I\over I_{\rm c}}d\lambda =
 \int (1-e^{-\tau(\lambda)})d\lambda
\label{w}
\end{equation}
where $I$ is the observed spectral intensity, $I_{\rm c}$ the interpolation
of the absorption-free continuum over the absorption feature and 
$\tau(\lambda )$ the optical depth. It is apparent from Eq.(\ref{w})
that the observed equivalent width does not depend on the spectral resolution.
For redshifted absorption lines, 
$w_{\rm obs}$~=~$w_{\rm rest}$$\times$(1+$z_{\rm abs}$).\par\noindent
If we assume that the absorbing atoms have a gaussian velocity distribution 
of mean $v_{\rm o}$ relative to the referential $R$, then 
it is possible to estimate the optical depth in Eq.(\ref{w})
as seen by an observer in $R$. An atom with velocity $v$ absorbs a photon of 
frequency $\nu$ in $R$ with a cross-section $\sigma$($\nu'$)
where $\nu'$~=~$\nu$/(1-$v/c$), $v$ taken to be positive for an atom that
goes away from the observer and,
\begin{equation}
\tau(\nu ) = N{1\over {\sqrt \pi} b}\int^{+\infty }_{-\infty } 
  \sigma(\nu')e^{-{(v-v_{\rm o})^2 \over b^2}} dv
\label{tau}
\end{equation}
where $N$ is the column density, the total number of atoms per surface unity
integrated over the gaussian distribution centered at $v_{\rm o}$
on the line of sight; 
$b$ is the Doppler parameter, $b$~=~$v_{\rm rms}$${\sqrt 2}$ = 
$FWHM$/2${\sqrt{Ln2}}$; $FWHM$ is the full width at half maximum.
When the velocity field is not turbulent, 
the Doppler parameter is related to the temperature of the gas 
by
\begin{equation}
b_{\rm th} = {\sqrt {2kT\over m}} = 12.8 {\rm (km\; s^{-1})} 
{\sqrt {T_4\over M}}
\label{bth}
\end{equation}
where $m$ is the mass of an atom, $M$ its mass number and
$T_4$~=~T/10$^4$~K. In case the turbulence can be modelled as
a gaussian velocity distribution with quadratic mean velocity
${\sqrt {3\over 2}} V$ and mean
velocity 2$V/{\sqrt \pi}$, then $b_{\rm tot}^2$~=~$b_{\rm th}^2$~+~$V^2$.
\par\noindent
The cross-section for the 
absorption in an atomic transition is characterized by an absolute
(classical)
value, modulated by the oscillator strength $f$ and a frequency dependence 
which is a consequence of the finite life-time of the upper level:
\begin{equation}
\sigma = f \times {1\over 4\pi\epsilon_{\rm o}} {\pi e^2\over m_{\rm e}c}
   \times {1\over \pi} {{\gamma\over 4\pi}\over (\nu-\nu_{\rm o})^2 + 
   ({\gamma \over 4\pi})^2}
\label{sigma}
\end{equation}
where the usual notations for the fundamental physical constants have
been used and $\gamma$ is the total damping constant (de-excitation
rate of the upper level). Eq.(\ref{tau}) reduces to
\begin{equation}
\tau(\lambda) = 1.498\times 10^{-2} {Nf\lambda\over b} H(a,u)
\label{tau2}
\end{equation}
with
\begin{equation}
H(a,u) = {a\over \pi}\int^{+\infty }_{-\infty }{e^{-y^2}\over (u-y)^2+a^2}
         dy;\; a = {\lambda\gamma\over 4\pi b};\; u = -{c\over b}\left(
         (1+{v\over c})-{\lambda\over \lambda_{\rm o}}\right)
\label{voigt}
\end{equation}
$H(a,u)$ is called the Voigt function which is a convolution of a
gaussian function and a lorentzian function. \par\noindent
The optical depth at the center of the line is:
\begin{equation}
\tau_{\rm o} = 1.497\times 10^{-15}{N{\rm (cm^{-2})}f\lambda_{\rm o}{\rm
       (\AA)} \over b{\rm (km\; s^{-1})}}
\label{tau0}
\end{equation}
More details can be found in e.g. \cite{mor73}.\par\noindent
%
%
\subsection{The Voigt Profile}
\begin{figure}
\centerline{
\vbox{
\psfig{figure=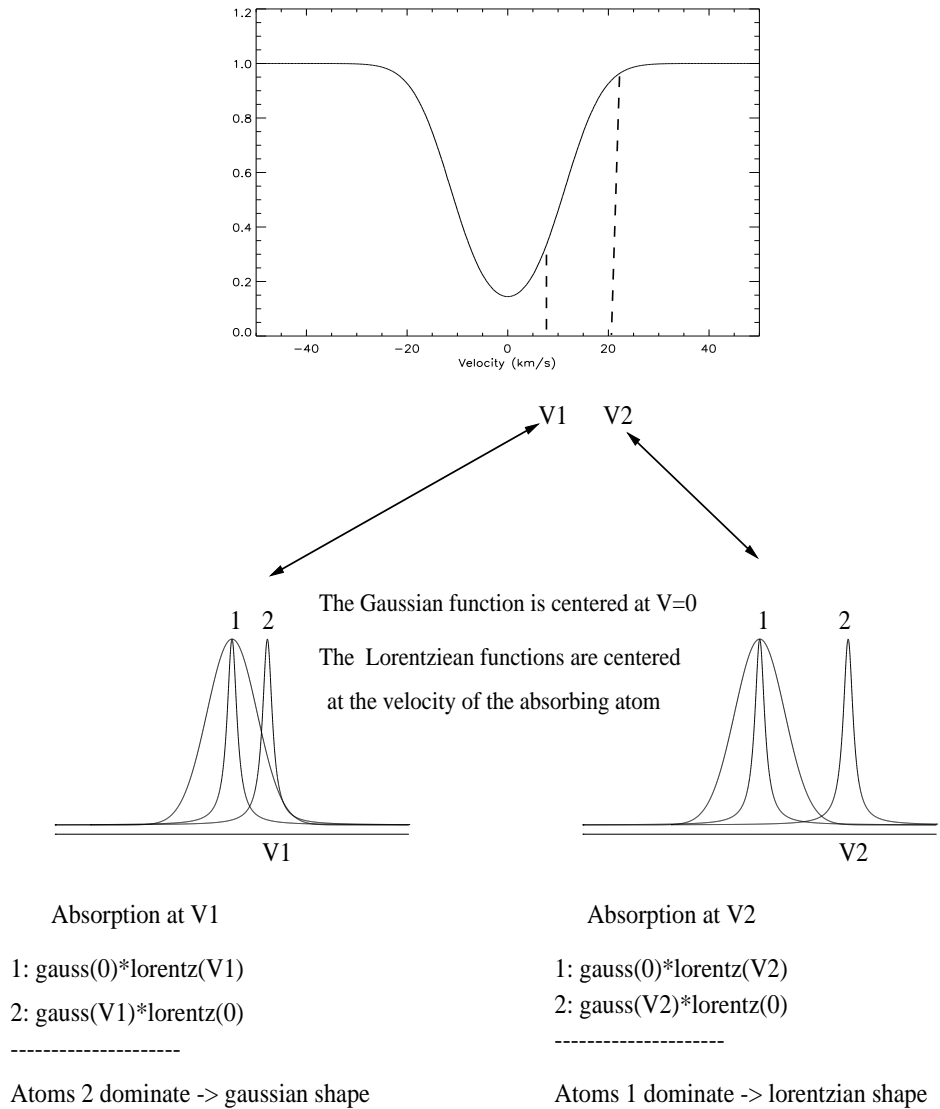,height=15.cm,width=13.cm,angle=0}
}}  
\caption{Illustration of the formation of the Voigt profile:
over the profile, the optical depth has a 
gaussian shape in the core of the absorption line and lorentzian 
shape in the extended wings (see the text). The functions have arbitrary
widths and are normalized to one at the maximum. 
}
\label{gl}
\end{figure}

Under the assumption that the velocity distribution of the atoms is
described by a Gaussian function, the overall shape of the
optical depth $\tau$ over the absorption line is a Voigt function.
The core of the Voigt function is gaussian and the far wings are
lorentzian. The absorption feature itself, which is simply $e^{-\tau}$, is 
called a Voigt profile. 
The different regimes are illustrated on Fig.~\ref{gl}. \par\noindent
The absorption of photons with wavelength $\lambda$1, or on a
velocity scale $V_1$~=~$c(\lambda_1-\lambda_0)/\lambda_0$, 
is due to a continuous sum of contributions by the
atoms with different velocities. The probability that atoms 1 of velocity
0 absorb the photons $\lambda_1$ is given by the value of the 
lorentzian at $V_1$. Since $V_1$ is close to the center of the 
absorption line, it is apparent that atoms 2 of velocity $V_1$ 
will dominate the absorption because the variations of the 
lorentzian is much faster than the variations of the gaussian from 
0 to $V_1$. The variations of the optical depth in the core of the
line will thus be governed by the gaussian function.
In contrast, the absorption of photons with wavelength $\lambda_2$, 
or on a velocity scale $V_2$~=~$c(\lambda_2-\lambda_0)/\lambda_0$,
is dominated by atoms with velocity 0. Indeed, far from the
center of the line, the variations of
the gaussian function are much faster than the variations of the lorentzian. 
The number of atoms with velocity $V_2$ is negligible. If absorption is
seen at $\lambda_2$, this is due to the non-negligible probability
of an atom 1 to absorb a photon with wavelength $\lambda_2$.
Of course, the total column density must be large
for the optical depth to be non negligible in the wing.
Thus when using the sketch of Fig.~\ref{gl}, it must be noticed 
that the transition between both regimes is very sharp around 
$u$~$\sim$~$\pm$3.2 (see Eq.\ref{voigt}) {\sl which corresponds to more than 
twice the FWHM}.
This is why the damping wings are prominent only for strong lines.
\par\noindent
Of course the assumption of a gaussian velocity distribution can be
questioned. Under other assumptions,
the absorption profile can be significantly different from
a Voigt profile. This is particularly important when discussing the 
characteristics of the absorptions derived from line profile fitting
(see \cite{kul95,lev97,lev98}).
\subsection{The Curve of Growth}
\begin{figure}
\begin{center}
\vskip -0.3cm
\centerline{
\vbox{
\psfig{figure=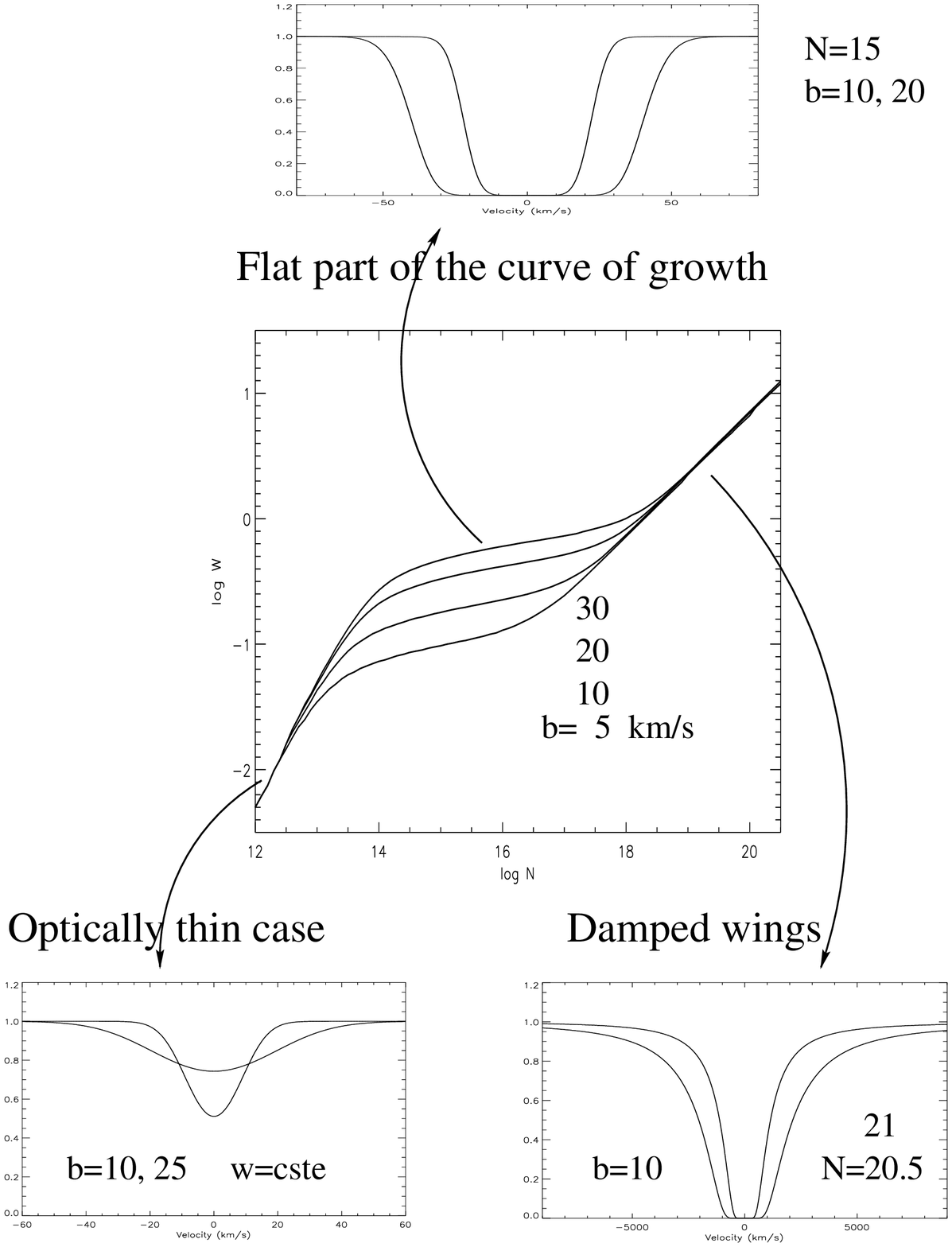,height=16.8cm,width=13.cm,angle=0}
}}  
\caption{Curve of growth: logarithm of the equivalent width ($w$ in
\AA) versus logarithm of the column density ($N$ in cm$^{-2}$) for 
different values of the Doppler parameter ($b$ in km~s$^{-1}$).
The curves are calculated for H~{\sc i} Ly$\alpha$1215. The three
characteristic regimes are illustrated (see the text).
}
\label{wn}
\end{center}
\end{figure}
From Eqs.(\ref{tau})(\ref{voigt})(\ref{w}) one can calculate the curve of 
growth which is the relation between the equivalent width of
a line and the column density for different values of the 
Doppler parameter. The curve of growth for the H~{\sc i}
Ly$\alpha$ transition is plotted on Fig.~\ref{wn} for $b$~=~5, 10, 20
and 30~km~s$^{-1}$. \par\noindent
There are three distinct regimes:\par\noindent
$\bullet$ When the column density is small ($\tau_{\rm o}$~$<$~0.1), 
the absorption line is 
optically thin and the equivalent width does not depend on $b$.
This is the linear part of the curve of growth, where the determination
of $N$ from $w$ is easy and reliable. For any transition,
\begin{equation}
N{\rm (cm^{-2})} = 1.13\times 10^{20}{w_{\rm r}{\rm (\AA)}\over
          \lambda^2{\rm (\AA)}f}
\label{nwthin}
\end{equation}
Fig.~\ref{wn} shows that for H~{\sc i}$\lambda$1215, a log~$N$~=~13 
line is partially saturated whatever the $b$ value is.\par\noindent
$\bullet$ The logarithmic or flat-part of the curve of growth 
is characterized by the large dependence of $N$ on $b$ at a given $w$.
In this regime, the determination of $b$ and $N$ are very uncertain 
except when several lines of the same ion are used. Equivalent
width and optical depth at the center of the line (see Eq.\ref{tau0})
are related by:
\begin{equation}
{w\over \lambda_{\rm o}} = 2 {b\over c} {\sqrt {Ln(\tau_{\rm o})}}
\label{nwflat}
\end{equation}
\par\noindent
$\bullet$ The absorption lines that are on the saturated part of the curve of 
growth are characterized by prominent damping wings. The equivalent
width does not depend on $b$ and the column density determination from
measurement of $w$ or fitting the wings is very accurate.
In that case (see Eqs.\ref{voigt} and \ref{tau0}),
\begin{equation}
{w\over \lambda_{\rm o}} = 2.64{b {\sqrt a}\over c}\times {\sqrt \tau_{\rm o}}
\label{nwdamp}
\end{equation}
For H~{\sc i}$\lambda$1215, $\gamma$~=~6.265$\times$10$^8$~s$^{-1}$
\cite{mor73} and:
\begin{equation}
N{\rm (cm^{-2})} = 1.88\times 10^{18} w_{\rm rest}^2{\rm (\AA)}
\label{ndamp}
\end{equation}
Fig.~\ref{wn} shows clearly that, for realistic values of $b$, the
H~{\sc i}$\lambda$1215 transition is in this regime for 
log~$N$~$>$~19. The definition of damped systems as absorbers with 
log~$N$(H~{\sc i})~$>$~20.3 is artificial and was introduced 
to search for damped candidates in low spectral resolution data
(see \cite{wol86}). Indeed,
for such column densities, the absorption line has 
$w_{\rm obs}$~$>$~15~\AA~ at $z$~$>$~2. The probability that such a
feature be the result of blending of weaker lines 
is small. For smaller column densities, the probability for such
confusion is much higher.\par\noindent
The values of $\lambda$ and $f$ are given for illustration
for a few transitions in Table~1. 
A more complete list can be found in \cite{mor88}\cite{vern94}\cite{sav96}.
\par\noindent
Note that most of the transitions are from the true ground state.
Exceptions are found (i) when the excited levels of the ground-state are
populated by pumping from the true ground-state by the background radiation 
(C~{\sc i}$^*$$\lambda\lambda$1560,1656, e.g. \cite{son94}, this leads
to determinations of the CMB temperature at high $z$;
C~{\sc ii}$^*$$\lambda$1334, e.g. \cite{les97}, is used to derive limits on
the electronic density) and (ii) in a few associated
or BAL absorption systems in which the density is high enough for
upper levels to be populated by collisional excitation (e.g. 
\cite{wam95}). \par\noindent
\begin{table}
\begin{center}
\begin{tabular}{l c l c c}
\multicolumn{5}{c}{Table 1. A few strong atomic transitions}\\
\\
\hline
\multicolumn{1}{c}{Ion}&\multicolumn{1}{c}{$\lambda_{\rm o}$}&
\multicolumn{1}{c}{f}&\multicolumn{1}{c}{log($\lambda_{\rm o}$f)}&
\multicolumn{1}{c}{log($\lambda_{\rm o}^2$f)}
\\
\multicolumn{1}{c}{}&\multicolumn{1}{c}{(\AA)}&
\multicolumn{1}{c}{}&\multicolumn{1}{c}{}&\multicolumn{1}{c}{}
\\
\hline
O~{\sc vi} & 1031.927 & 0.130  & 2.128 & 5.141 \\ 
O~{\sc vi} & 1037.616 & 0.0648 & 1.828 & 4.844 \\
H~{\sc i}  & 1215.670 & 0.4162 & 2.704 & 5.789 \\
O~{\sc i}  & 1302.169 & 0.0486 & 1.801 & 4.916 \\
C~{\sc ii} & 1334.532 & 0.118  & 2.197 & 5.323 \\
Si~{\sc iv}& 1393.755 & 0.528  & 2.867 & 6.011 \\
Si~{\sc iv}& 1402.770 & 0.262  & 2.565 & 5.712 \\
C~{\sc iv} & 1548.202 & 0.194  & 2.448 & 5.667 \\
C~{\sc iv} & 1550.774 & 0.097  & 2.177 & 5.368 \\
Mg~{\sc ii} & 2796.352 & 0.592 & 3.219 & 6.666 \\
Mg~{\sc ii} & 2803.531 & 0.295 & 2.918 & 6.365 \\
\hline
\end{tabular}
\end{center}
\end{table}
%
%
The column density and Doppler parameter for an ion can be formaly 
determined if two transitions from the same lower level are observed
on the flat part of the curve of growth.
Indeed from Eqs.(\ref{nwflat}) and (\ref{tau0}) it is easy to extract 
$N$ and $b$. This is the doublet method which has been
extensively used because doublets are numerous and easily observed
(see Table~1). Of course the parameters are better determined when more 
than two lines from the same level are used. \par\noindent
\subsection{Fitting the Lines}
\begin{figure}
\vskip -1.3cm
\centerline{
\vbox{
\psfig{figure=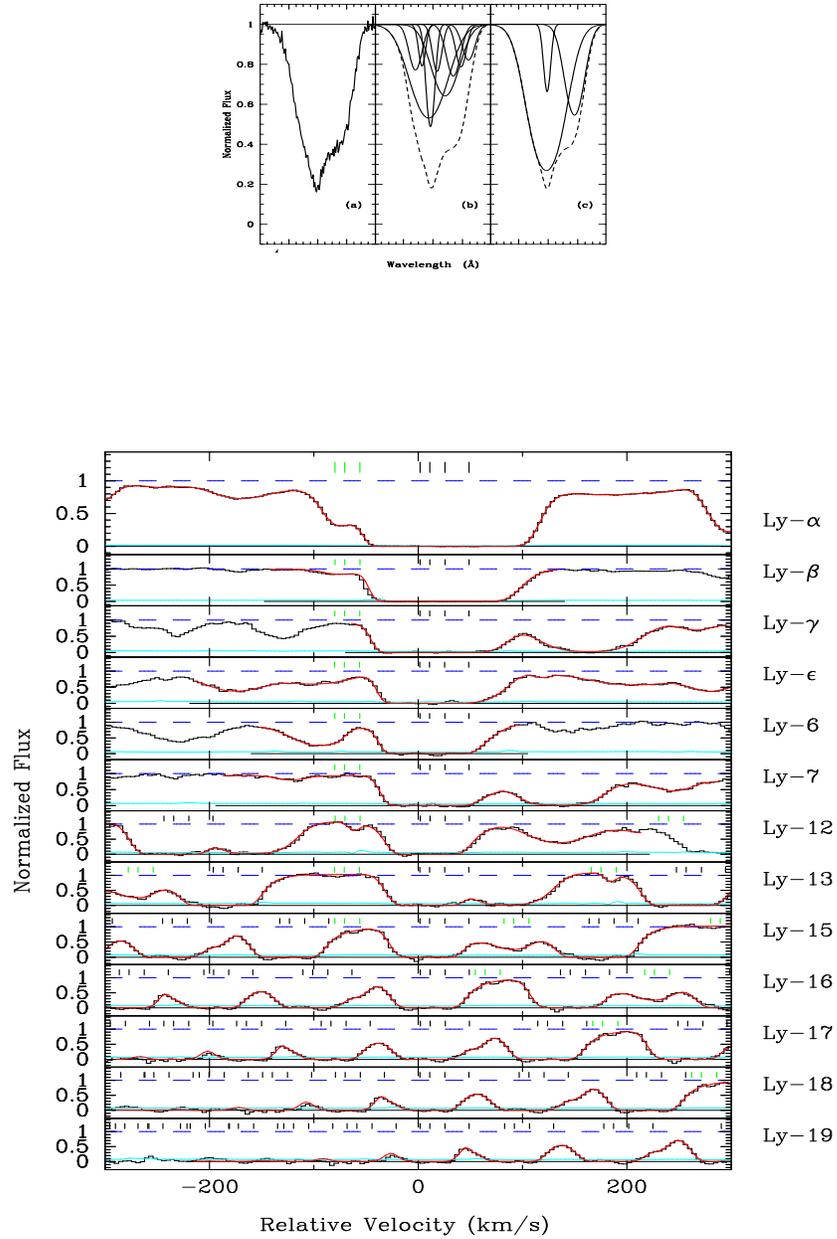,height=21.cm,width=13.cm,angle=0}
}}  
\vskip -2.5cm
\caption{Upper panel: The fit of a single absorption line (here 
H~{\sc i}$\lambda$1215; from Fernandez-Soto et al. \cite{fer96}) is by 
no means unique; to constrain the model,
several transitions are used, either from the same ion (in the lower panel
all the H~{\sc i} transitions in the Lyman series are used; from 
Burles \& Tytler
\cite{bur98}; note that the model is overplotted to the data) or from
different ions assumed to arise in the same phase.
}
\label{fit}
\end{figure}

\begin{figure}
\centerline{
\vbox{
\psfig{figure=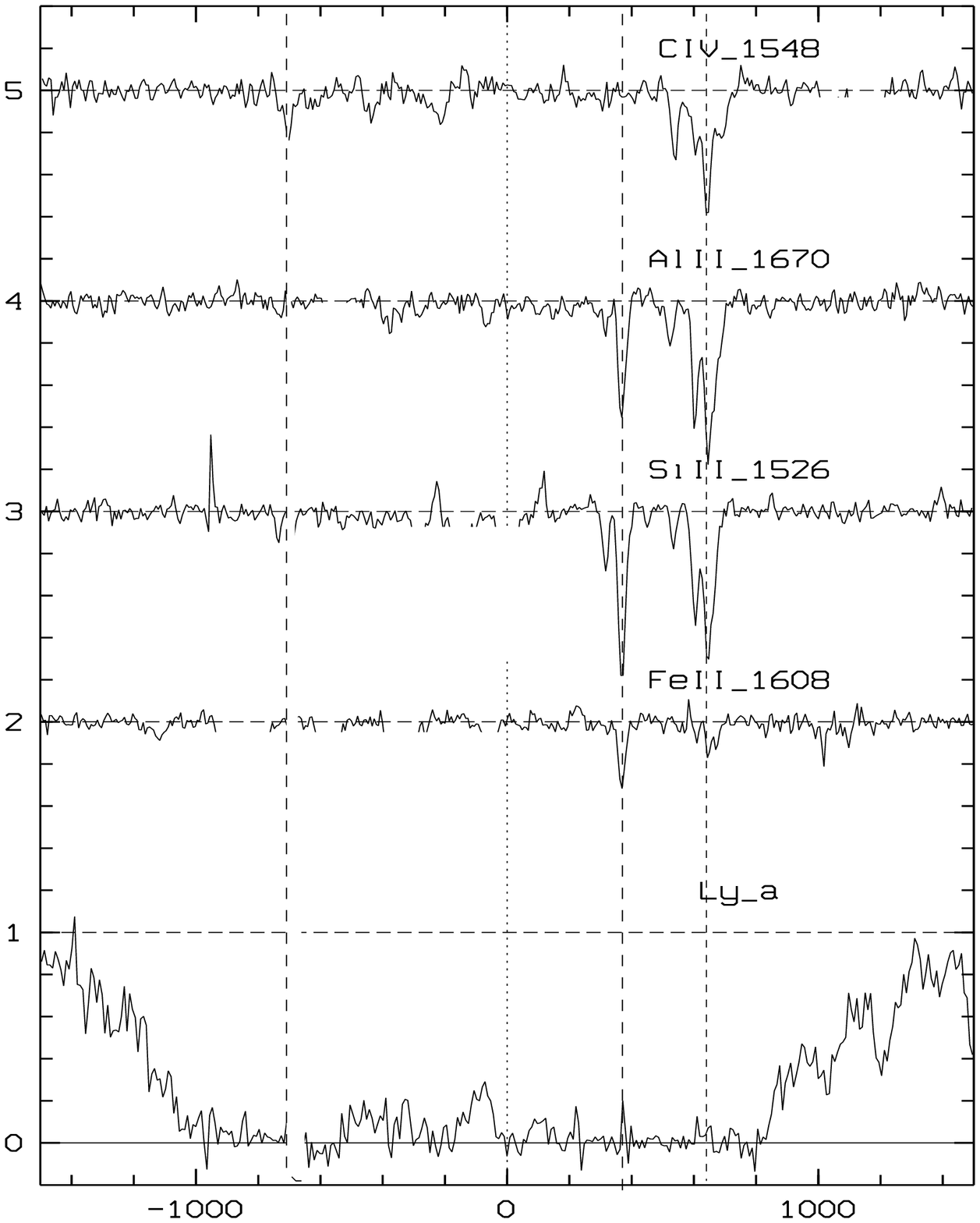,height=12.cm,width=10.5cm,angle=0}
}}  
\vskip -0.5cm
\caption{Absorption lines in the $z_{\rm abs}$~=~3.1825 system
toward PKS~2000--330 plotted on a relative velocity scale.
Parts of the spectrum have been removed when absorption lines 
from other systems were present. It is apparent that 
(i) the metal lines are much narrower than the H~{\sc i} line; note
however that metallicity in the system is quite low, of the order
of 1/100 of solar; 
(ii) the number of components needed to fit the H~{\sc i} line is much
larger than for the metal lines;
(iii) the position of three metal components have been marked by dashed
vertical lines; it can be seen that the velocity coincidence is very good 
between Si~{\sc ii}, Al~{\sc ii} and Fe~{\sc ii}.
}
\label{q2000}
\end{figure}

In practice however the problem is complicated by the structure of
the absorber. Most often, an absorption feature is a blend of
several components. Indeed when the line of sight passes for example
through a galactic halo, it can intercept several clouds. The relative 
projected velocities of the clouds are usually not very large 
(typically $\Delta V$~$<$~150~km~s$^{-1}$) compared to the width of the 
lines (5 to 50~km~s$^{-1}$ or larger) and the number of clouds can be
large (more than ten). Column densities and 
Doppler parameters for each component are derived from line profile
fitting, using the maximum of transitions.
This is illustrated on Fig.~\ref{fit} for H~{\sc i}. The fit of a single 
absorption feature with several components is most often not unique; 
the different 
solutions can yield column densities differing by an order of magnitude. 
When several transitions from the same ion are used, as the transitions
in the Lyman series for H~{\sc i}, better significance is achieved. 
Stronger constraints can be put on models by using several ions. Indeed  
metal lines are narrower than H~{\sc i} lines and column densities are 
usually smaller for metals than for H~{\sc i}, so that metal absorption
features are relatively less sensitive to blending effects
(see Fig.~\ref{q2000}). However it must be assumed that the 
decomposition in components is the same for the different ions.
Hence, it is usually a good assumption to consider
that H~{\sc i}, O~{\sc i}, Si~{\sc ii}, C~{\sc ii}, Al~{\sc ii}
and Fe~{\sc ii} on the one hand, and C~{\sc iv} and Si~{\sc iv} on the other 
arise in the same phase and can be fitted together
(see Fig.~\ref{q2000}).
\par\noindent
Several procedures using $\chi^2$ minimization techniques are available 
to perform Voigt profile fitting of the absorption features.
The most popular are FITLYMAN implemented in MIDAS the ESO image processing 
package (Fontana \& Ballester \cite{fon95}) and VPFIT  (Carswell et al. 
\cite{cars92}). To visualize the lines XVOIGT is also very convenient 
(Mar \cite{mar94}). Most of the models overplotted on the figures have
been created with the latter procedure. 
\par\noindent
\section{THE LY$\alpha$ FOREST}
The numerous H~{\sc i} lines seen in the Ly$\alpha$ forest correspond
to the imprint left by the intergalactic medium in the QSO spectrum.
The whole redshift interval from zero to the highest QSO emission 
redshift has been observed. 
A review of the recent works and ideas on this subject can be found in 
\cite{rau98}. In the following, a few important observations
and interpretations are highlighted.\par\noindent
\subsection{Evolution of the Number Density of Lines}
\begin{figure}
\vskip 1.3cm
\centerline{
\psfig{file=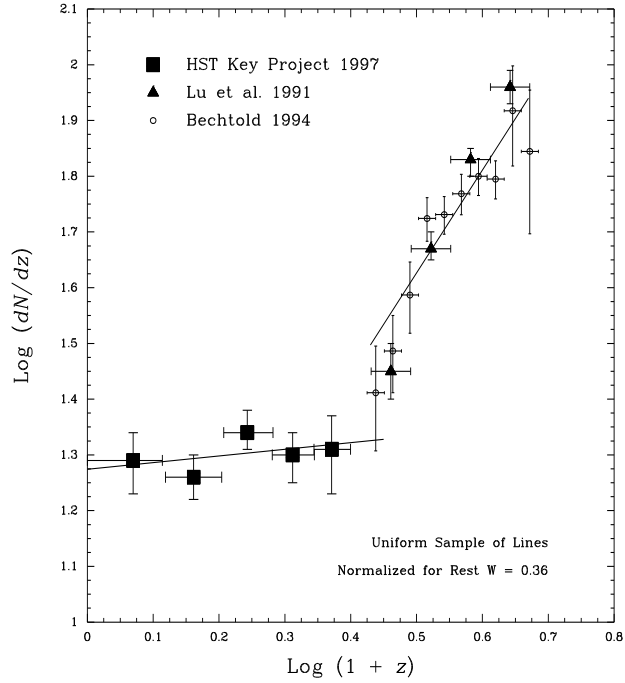,height=7.cm,width=10cm,angle=0}
}
\caption{Logarithm of the number density of Ly$\alpha$ lines 
versus redshift from Jannuzi et al. \cite{jan98}. The break at
$z$~$\sim$~1.5 is apparent.}
\label{dndzlya}
\end{figure}

The redshift evolution of the number of Ly$\alpha$ lines per unit redshift 
is usually fitted by a power law,
\begin{equation} 
N(z)=N_{\rm o}(1+z)^{\gamma} 
\label{nz} 
\end{equation}
Note that $N_{\rm o}$ depends on the equivalent width; weak lines
are more numerous than strong lines (see next Section).
The observed exponent $\gamma$ can be compared with that expected 
if the forest is due to a non-evolving population: 
\begin{equation}
{\gamma}=2-{1\over 2}[3\Omega_{\rm o}(1+z)^3-2K_{\rm o}(1+z)^2]
\times[\Omega_{\rm o}(1+z)^3-K_{\rm o}(1+z)^2+{\Lambda \over 3H_{\rm
o}^2}]^{-1}
\label{gammatot}
\end{equation}
with standard notations and with 
$K_{\rm o}$~=~$\Omega_{\rm o}$+${\Lambda \over 3H_{\rm o}^2}$--1 (see
\cite{fuk91}). For a model without a cosmological constant,
\begin{equation}
{\gamma}=(1+q_{\rm o}z-q_{\rm o})(1+2q_{\rm o}z)^{-1}
\label{gamma} 
\end{equation}
In the absence of evolution, the exponent is equal to 1.0 for 
$q_{\rm o}$~=~0 and 0.5 for $q_{\rm o}$~=~0.5 at $z$~=~2. 
Observations show that the Ly$\alpha$ forest evolves strongly
at high redshift (\cite{kim97}; $\gamma$~$\sim$~2.5 at $z$~$\sim$~2);
that weak lines evolve more slowly than stronger one; and that
there is a break at $z$~$\sim$~1.5 in the cosmological evolution
of the number density of lines (see Fig.~\ref{dndzlya}, \cite{jan98}).
\subsection{Column Density Distribution}
\begin{figure}
\centerline{
\psfig{file=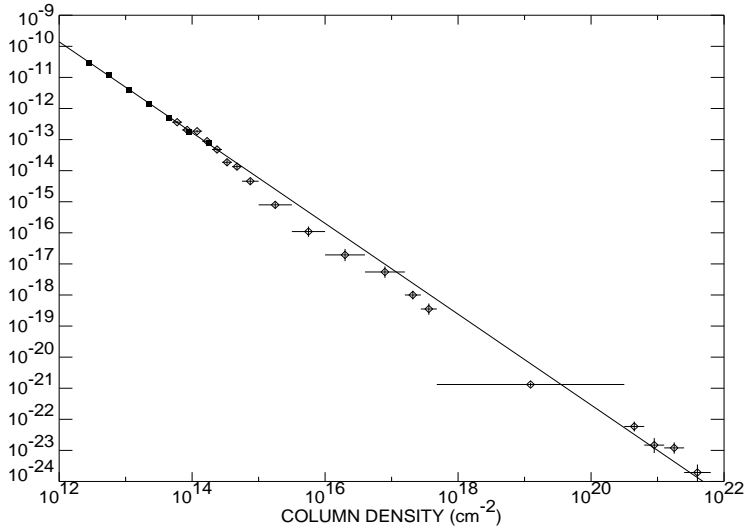,height=7.cm,width=10.5cm,angle=0}
}
\caption{H~{\sc i} column density distribution at $z$~$\sim$~2.8
from Hu et al. \cite{hu95} and Petitjean et al. \cite{pet93}. There is a
deficit of lines at $N$~$\sim$~10$^{15}$~cm$^{-2}$.
}
\label{disth1}
\end{figure}

The column density distribution for all absorption systems
(the Ly$\alpha$ forest corresponds only to the smallest 
column densities) at $z$~$\sim$~2.8 is shown on Fig.~\ref{disth1}.
The best fit for the forest is a power-law 
d$^2n$/d$N$d$z$~$\propto$~$N^{-1.5}$. There is a deficit of lines
at $N$~$\sim$~10$^{15}$~cm$^{-2}$ and a flatening of
the distribution for the damped systems \cite{kim97, pet93}. 
These features have been interpreted as a consequence of the 
structure of the intergalactic medium and of different 
ionization effects \cite{char94, far98, pet92}.
It has been shown from this distribution that the Ly$\alpha$ forest
at high redshift contains most of the baryons in the Universe
\cite{pet93}, a result that has been confirmed by
other methods \cite{pres93, mei93}. This suggests that at high redshift,
the intergalactic medium is the reservoir of baryons for galaxy formation.
\subsection{Physical Conditions}
The density of the gas must be very small. Indeed from coincident
absorptions along adjacent lines of sight (see below, e.g.
\cite{bech94, din98}),
it can be shown that the gas {\sl in the Ly$\alpha$ forest} is fairly 
uniform on scales of $\sim$1~kpc. This implies that for a typical column
density of $N$(H~{\sc i})~=~10$^{14}$~cm$^{-2}$, the H~{\sc i} density is
smaller than 3$\times$10$^{-8}$~cm$^{-3}$. Even for an ionization correction 
factor corresponding to highly ionized gas, $n$(HII)/$n$(HI)~=~10$^{4}$,
the total density is smaller than 3$\times$10$^{-4}$~cm$^{-3}$.\par\noindent
The intergalactic gas is exposed to the effect of the UV-background
which is the radiation field produced by QSOs and early galaxies at high
redshift. It is possible to estimate the flux at the Lyman limit
using the observed number of sources (e.g. \cite{mir90, rau97}),
$J_{912}$~$>$~2.3$\times$10$^{-22}$~erg/s/cm$^{2}$/Hz/sr at $z$~$\sim$~2.5
if the shape of the ionizing spectrum is taken as in \cite{haa96}.
With the above orders of magnitude the ionization parameter\footnote{Ratio 
of the density of ionizing photons to the total hydrogen density;
since $n_{\rm HI}\times\sigma F/h\nu \propto n_{\rm e}^2\times
\alpha$, with $\sigma$ the ionization cross-section, $\alpha$ the 
recombination coefficient, $n_{\rm HI,HII,e}$ the HI, HII and electronic
densities and $F$ the ionizing flux, then
$n_{\rm HII}/n_{\rm HI}\propto \sigma/\alpha\times cU$}
is $U$~=~0.02 and the gas is indeed completely ionized.\par\noindent
The presence of metals in the Ly$\alpha$ forest is an important clue
towards understanding how and where the first objects in the universe
formed. It has been known for long that the amount of metals is 
small. However given the small column densities involved, the question
could not be satisfactorily answered before the advent of the 
10m-class telescopes. Using Keck data, it has been shown recently
that C~{\sc iv} absorption is seen in all the systems with 
log~$N$(H~{\sc i})~$>$~15 and in half of the systems with 
log~$N$(H~{\sc i})~$>$~14.3 \cite{cow95, tyt95, son96}. 
A lower limit on the metallicity,
$Z$~$>$~10$^{-3}$~$Z_{\odot}$ ($H_{\rm o}$~=~65, $q_{\rm o}$~=~0.02),
is derived after difficult ionization corrections \cite{son98}.
The production of this amount of metals would induce enough photons 
to completely pre-ionized the IGM \cite{mir97}. However, it is
much lower than the predictions of metal production by a first generation of
Type II supernovae \cite{mir97}.
The presence of metals in systems with smaller column densities 
is still controversial \cite{cow98, lu98}.\par\noindent
Another way to estimate the ionizing UV-background is to use the 
proximity effect which is the observed decrease of the number
of absorption lines at a given equivalent width limit in the vicinity of 
the QSO. This is interpreted as a consequence of the enhanced  
ionizing flux due to the emission by the quasar.
The radius of influence of the quasar emission can thus be derived from
observations. This is the distance
from the QSO at which the depleted number density of lines reaches
to the background value. At this point, to a first approximation, the
flux from the background is comparable to the flux from the quasar
\cite{baj88}. However this determination 
assumes that (i) the number density of clouds can be extrapolated
toward the quasar (this may be wrong if the density of clouds increases
in the potential wells of the QSO), (ii) the luminosity of the QSO is
known from direct observation (questionable if variability occurs), 
(iii) the redshift of the quasar is known (discrepancies between different
lines are the rule, \cite{esp93}). From this it is easy to understand that
the typical value of $J_{912}$~$\sim$~10$^{-21}$~erg/s/cm$^{2}$/Hz/sr
derived from these studies is highly uncertain. The uncertainties
are discussed in \cite{bech94b, gia98} and in more detail in
\cite{coo97}.
\subsection{Clustering Properties}
\begin{figure}
\centerline{
\psfig{file=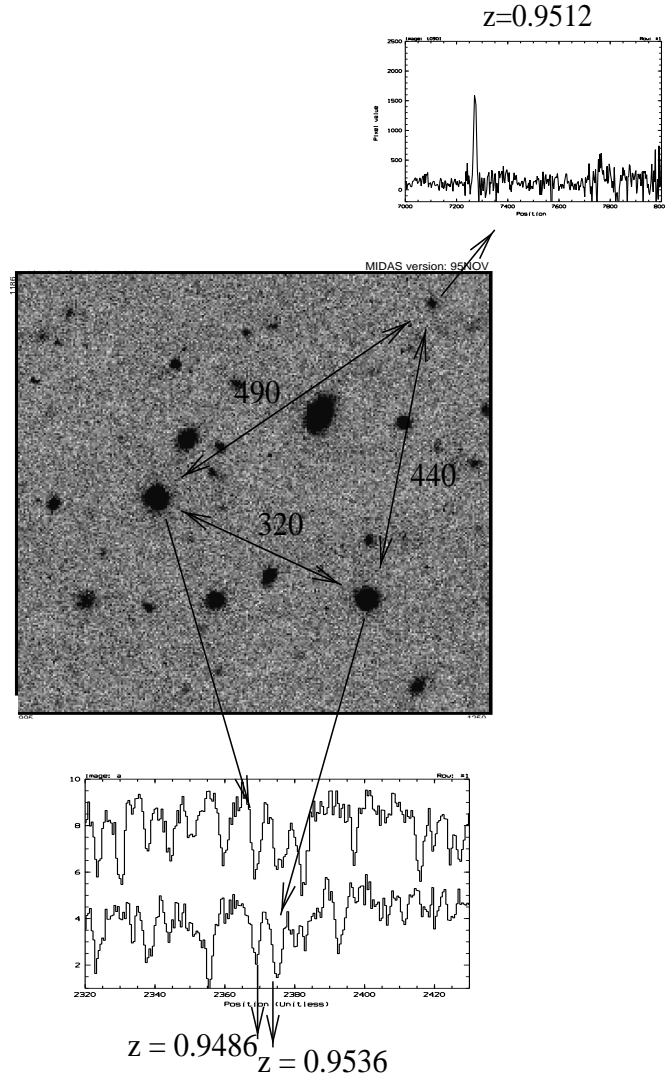,height=14.5cm,width=9.cm,angle=0}
}
\caption{Observation of a quasar pair \cite{pet98}: 
Q1026--0025A ($z_{\rm em}$~=~1.438; situated in the bottom right corner
of the image) and B ($z_{\rm em}$~=~1.526; 36~arcsec or 320$h^{-1}_{50}$~kpc 
away from A in the north-east direction). North is at the top, east to 
the left. Ly$\alpha$ absorptions are seen at $z_{\rm abs}$~=~0.9486 and 
0.9536 in the HST spectrum of A and at $z_{\rm abs}$~=~0.9486 in B.
A galaxy is detected with the NTT (ESO-La Silla) at redshift 
$z_{\rm em}$~=~0.9512 at a projected distance of $\sim$~450~kpc from the
two absorbers. Using 10m-class telescopes, it will be possible to 
observe $m$~$\sim$~22 magnitude quasar at intermediate resolution.
The aim is to increase the density of lines of sight in order to study the 3D 
spatial distribution of the gas and to correlate this distribution with
the galaxies in the field.}
\label{pairs}
\end{figure}

The two-point correlation function $\xi$\footnote{The probability to find 
a Ly$\alpha$ cloud in a volume d$V$ at a distance $r$ from another cloud is
given by d$P$~=~$\Phi$($z$)d$V$[1~+~$\xi$($r$)], where $\Phi$ is the
spatial density of clouds.} has been studied by several authors 
with contradictory conclusions, probably because the number of lines of
sight was not large enough. Cristiani et al. \cite{cri97} use 
1600 lines and show that there is a strong signal on scales 
$\Delta v$~$<$~200~km~s$^{-1}$ which increases for increasing column
densities. However, it is difficult to disentangle the large scale
clustering signal from the effects of  
the internal structure in an absorber \cite{fer96}. Indeed, the
column density through a system increases because the number of 
subcomponents increases;
each component beeing the signature of an individual cloud in an extended halo
\cite{petb94}. This is why observers have tried to measure directly the 
dimensions of absorbing clouds. For this, they use pairs or groups of quasars 
with small angular separations on the sky. Correlation of absorption
features along the different lines of sight indicates that the characteristic
size of the absorbers is larger than the projected separation of the
lines of sight.
In the spectra of multiple images of lensed quasars with separations
of the order of a few arcsec \cite{sme95, imp96}, the Ly$\alpha$
forests appear nearly identical, implying that the absorbing objects
have large sizes ($>$50~kpc). For larger separations between lines of 
sight, 
the correlation between the two forests decreases; however
after random coincidences are removed, an excess of common redshift 
absorption exists even for separations as large as 500~$h^{-1}$~kpc
\cite{cro98}. This suggests very large dimensions or at least 
that the clouds are closely correlated on these scales.
The data available up to now have been critically
analyzed by D'Odorico et al. \cite{dod98} who concluded that 
the absorbers have typical sizes of the order of
$R$~$\sim$~350$\pm$100$h^{-1}$~kpc. The number of suitable groups of quasars 
is too small at present to constrain the structure of the Ly$\alpha$ complexes.
However, it is easy to see that the method is very powerful to study 
the correlation of baryonic matter at high redshift. The number 
of observable pairs should increase dramatically with the advent of
10m-class telescopes. It is also conceivable, although ambitious, to
probe the 3D-distribution of the gas in a small field by increasing the 
density of lines of sight. To this aim, quasars with magnitude 22 should 
be observed at intermediate resolution. The large amount of telescope time
needed requires multi-object spectroscopy capabilities on 10m-class telescopes
\cite{pet97}. If a hundred lines of sight could be observed in one square 
degree, the probability distribution function of the matter density could be 
investigated at high redshift. This would be a unique way to 
study how galaxy formation is related to the distribution and dynamics of the
underlying matter field (see \cite{nus98, crof98} and next Section). 
The method is illustrated in Fig.~\ref{pairs}. 
\par\noindent
\subsection{Simulations}
Most models of the Ly$\alpha$ forest assume that the 
intergalactic medium is populated by diffuse clouds, either gravitationally or
pressure confined, and possibly embedded in a pervasive medium 
(e.g. \cite{bla81, sar80, ree86, ike86}).
A more attractive picture arised recently from large-scale dark-matter
$N$-body simulations including a description of the baryonic component:
either full hydrodynamic calculations 
\cite{cen94, her96, mir96, zha98}
limited however to $z$~$\sim$~2, or an approximate treatment that 
can follow evolution down to $z$~=~0
\cite{pet95, muc96, rie98, bi97}. The simulations show
that, to a very good approximation, 
the properties of the Ly$\alpha$ forest can be understood if the
gas traces the gravitational potential of the dark matter.
In this picture, part of the gas is located inside filaments
where star formation can occur very early in small haloes that subsequently
merge to build up galaxies. 
It is thus not surprizing to observe metal lines from this gas.
The remaining part of the gas
either is loosely associated with the filaments and 
has $N$(H~{\sc i})~$>$~10$^{14}$~cm$^{-2}$,
or is located in the underdense regions and has 
$N$(H~{\sc i})~$<$~10$^{14}$~cm$^{-2}$.
Much of the Ly$\alpha$ forest arises in this gas and
it is still to be demonstrated that it contains metals \cite{cow98, lu98}.
To clarify these issues, it is important to follow the evolution of the 
Ly$\alpha$ gas over a wide redshift interval down to the 
current epoch ($z$~=~0), distinguishing between gas 
closely associated with the filamentary structures of the 
dark matter and the dilute gas mainly located in the underdense regions.
\par
\begin{figure}
\centerline{
\psfig{file=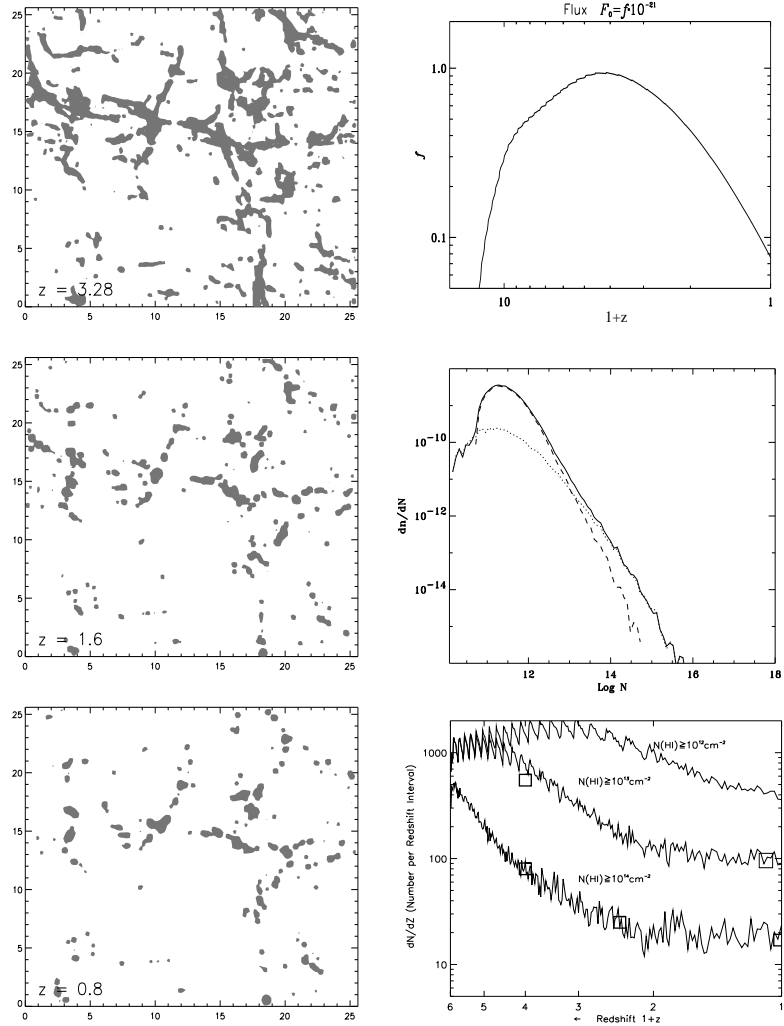,height=14.cm,width=11cm,angle=0}
}
\caption{Left handside panels: 25$\times$25$\times$2~Mpc$^3$ slice through the 
simulation box at $z$~=~3.28, 1.6 and 0.8; contours delineate
regions where
$N$(H~{\sc i})~$>$~10$^{14}$~cm$^{-2}$.  Right handside panels: Evolution
of the ionizing UV-background flux (calculated from the amount of gas
cooling below 5000~K in the simulation); H~{\sc i} column density
distribution at $z$~$\sim$~2;
number density of lines with $N$(H~{\sc i})~$>$~10$^{12}$,
10$^{13}$ and 10$^{14}$~cm$^{-2}$ versus redshift (observational points are
overplotted as squares; note the low redshift points from \cite{tri98}).
The simulation is from Riediger et al. \cite{rie98}.}
\label{simul}
\end{figure}

\noindent
The simulations are successful at reproducing the global properties
of the Ly$\alpha$ forest and its evolution (see Fig.~\ref{simul}).
They have favored the emergence of a consistent picture in which
not only the intergalactic medium is the baryonic reservoir for galaxy
formation, but also galaxy formation strongly influences the evolution
of the IGM through metal enrichment and ionizing radiation emission.
More information on the 3D spatial distribution of the gas should be
gathered from observation of multiple
lines of sight to add constraints to these models (see previous Section).
\section{METAL LINE SYSTEMS AND GALAXIES}
Detailed studies of the kinematics and metallicities of metal line systems
(see e.g. \cite{ste90a, ste90b, pet90, petb94}) are unique for
understanding the structure and evolution of galactic haloes. 
Indeed the connection between apparently normal galaxies and strong metal 
absorption line systems has been explored in some detail recently
\cite{ber91, ste94, gui97}. The fields of quasars known to have 
metal absorption line systems at $z_{\rm abs}$ are searched for galaxies at
redshift $z_{\rm em}$~$\sim$~$z_{\rm abs}$. Up to
$z$~$\sim$~1, there is always such a galaxy within $\sim$~10~arcsec
from the line of sight to the quasar.\par\noindent
At intermediate redshift ($z$~$\sim$~0.7) Mg~{\sc ii} systems with 
equivalent width $w_{\rm r}$ $>$ 0.3~\AA~ are primarily associated
with  large $R$~$\sim$~35$h^{-1}$~kpc haloes of a population of luminous 
($\sim$$L_{\rm B}^*$) galaxies which appear not to evolve 
significantly over the observed redshift range. 
The origin, structure and evolution of these large galactic haloes
play an important role for the formation and evolution of galaxies, 
as there is a close interaction between the disk where 
most of the stars form  and the halo that acts as a reservoir
of gas. \par
\begin{figure}
\vskip -5.cm
\centerline{\hskip -1cm
\psfig{file=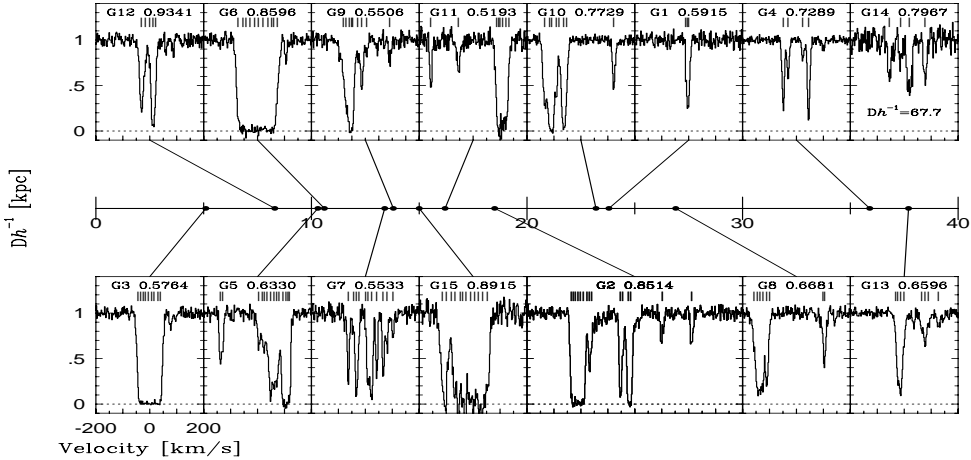,height=19.cm,width=20cm,angle=0}
}
\vskip -7.cm
\caption{The Mg~{\sc ii} absorption line profiles for ten systems are
plotted versus the impact parameter of the field-galaxy 
closest to the line of sight at the same redshift as the absorber.
It is apparent that there is no correlation between the absorption profile and
the impact parameter. Although this comparison is very crude
since the absorbing galaxies have neither the same luminosity nor
the same star-formation rate, this nonetheless suggests that galactic haloes 
have highly perturbed structures (from Churchill et al. \cite{chu96}).}
\label{mg2}
\end{figure}

\noindent
The most striking result of these studies is that the only 
interlopers (galaxies that are close to the line of sight but do not 
induce absorption in the QSO spectrum) have 
$L_{\rm K}$~$<$~0.05~$L^*_{\rm K}$. At intermediate 
redshift, the most important criterion for selecting absorbing galaxies
seems therefore to be  mass rather than star-formation rate.
However, it has been shown that  
the spectral energy distribution of absorbing galaxies is  bluer at
higher redshift and that the mean equivalent
width of [OII]$\lambda$3727 increases by 40\% between $z$~$\sim$~0.4 and 
$z$~$\sim$~0.9 \cite{gui97}. This suggests that star-formation activity is  
also an important  factor for the presence of large haloes.
In any case it seems that the structure of haloes is highly perturbed.
In particular, it has been shown that the kinematics of the absorption 
systems is {\sl not} correlated with the impact parameter (\cite{chu96}
and Fig.~\ref{mg2}).
Simple disk or halo models are rejected \cite{char98}.\par\noindent
It is also still unclear which galaxies are associated with weak 
($w_{\rm r}$(Mg~{\sc ii})~$<$~0.3~\AA) metal absorption systems. 
The characteristic radius of haloes can be estimated from the 
incidence rate of absorption, assuming that every galaxy is surrounded
by a spherical halo with filling factor unity, and that the relation
between luminosity and halo radius derived from Mg~{\sc ii} galaxy detection
can be extrapolated to small equivalent width. The detection limit for
weak metal lines has been pushed to very low values with  
10m-class telescopes. Mg~{\sc ii} systems are expected to trace the
gas with highest density. The number  of 
Mg~{\sc ii} systems per unit redshift at 0.4~$<$~$z$~$<$~1.4 with 
$W_{\rm r}$~$>$~0.02~\AA~ is d$N$/d$z$~=~1.74$\pm$0.10 \cite{chu98}.
This is four times larger than the number of Lyman limit systems and 
corresponds to 5\% of the number of Ly$\alpha$ forest lines with 
$W_{\rm r}$~$>$~0.1~\AA. This implies a halo radius of the order of 
63$h^{-1}$~kpc for $L^*$ galaxies. This suggests that 
either normal galaxies are surrounded by huge haloes or  
that weak systems are associated with a population of dwarf galaxies.
\par\noindent
\section{DAMPED SYSTEMS}
Damped Ly$\alpha$ (hereafter DLA) systems are defined as systems with
hydrogen column density $N$(H~{\sc i})~$>$~2$\times$10$^{20}$~cm$^{-2}$. 
This definition is artificial since damped wings appear for
lower column densities ($N$(H~{\sc i})~$>$~10$^{19}$~cm$^{-2}$;
see Section 2.3 and Fig.~\ref{wn}). It has been 
introduced assuming that these lines should be characteristic of 
galactic disks at high redshift \cite{wol86}.
Another reason is that damped Ly$\alpha$ surveys were performed
at low resolution. The equivalent width 
$w_{\rm obs}$($z$~$\sim$~2.5)~$>$~17.5~\AA~ for 
$N$(H~{\sc i})~$>$~10$^{20}$~cm$^{-2}$. The probability that such a 
strong absorption feature is the result of blending is small
(see Table~3 of \cite{wol86}). It is clear that this definition may 
introduce a systematic bias in the discussion of what is the nature of
these systems and it would be most valuable to
compare the properties of systems with 
19~$<$~log~$N$(H~{\sc i})~$<$~20.3 and log~$N$(H~{\sc i})~$>$~20.3.
\par\noindent
For log~$N$(H~{\sc i})~$>$~19, the optical depth 
at the Lyman limit is large enough so that hydrogen is neutral. 
The gas is either cold
($T$~$<$~1000~K) and contains molecules (e.g. \cite{sri98})
or warm ($T$~$\sim$~10$^4$~K) for the highest or lowest
column densities respectively \cite{pet92}. 
As a consequence of the shape of the column density distribution, 
d$^2$$n$/d$N$d$z$~$\propto$~$N^{-\beta}$ with $\beta$~$\sim$~1.5, 
most of the mass is in the systems of highest column densities.
The number density of the latter decreases with time presumably as a
consequence of star-formation (\cite{wol86, lanz95}; 
see however \cite{turn98}). Indeed, the cosmic 
density of neutral hydrogen in
DLA absorbers at $z$~$\sim$~3 is similar to that of stars at the present 
time \cite{wol95, stor96}.\par
\begin{figure}
\vskip -5cm
\centerline{
\psfig{file=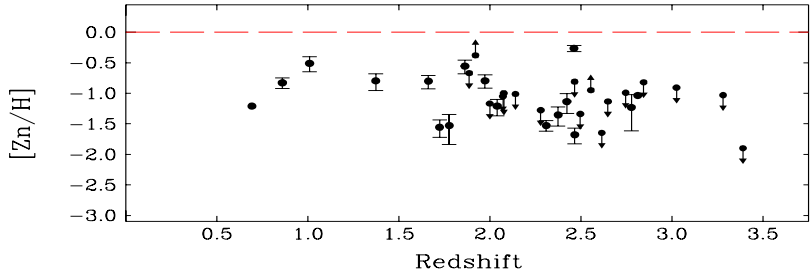,height=17.cm,width=15cm,angle=0}
}
\vskip -7cm
\caption{Zinc metallicity (relative to solar) in damped systems versus
redshift (from Pettini et al. \cite{pett97b}).}
\label{abdamp}
\end{figure}

\noindent
Metallicities and dust content
have been derived from zinc and chromium observations 
\cite{pett94, pett97a}. The [Zn/Cr] ratio
is an indicator of the presence of dust if it is assumed,
that, as in our Galaxy, zinc traces the gaseous abundances 
whereas chromium is heavily depleted into dust grains
(see \cite{lu96} and \cite{pett97b} for a critical discussion of this
assumption). The typical dust-to-gas ratio determined this way 
is about 1/30 of that in the Milky-Way \cite{pett97a, vlad98}. This
amount of dust could bias the observed number density of DLA systems
\cite{fall93, bois98}.
Metallicities are of the order of a tenth solar and tend to decrease 
from $z$~$\sim$~2 to $z$~$>$~3 (\cite{pett97b, bois98} and Fig.~\ref{abdamp}).
At any redshift however, the scatter is large and it may be hazardous to draw
premature conclusions from the small samples available. \par\noindent
Recently, Prochaska \& Wolfe \cite{proc97} have used Keck spectra of 17 DLA
absorbers to investigate the kinematics of the neutral gas from unsaturated
low-excitation transitions such as Si~{\sc ii}$\lambda$1808. They show that
the absorption profiles are inconsistent with models of galactic haloes with
random motions, spherically infalling gas and slowly rotating hot disks. The
CDM model \cite{kauf96} is rejected as it produces disks with
rotation velocities too small to account for the large observed velocity 
broadening of the absorption lines.
Models of thick disks ($h$~$\sim$~0.3$R$, where $h$ is the vertical scale and
$R$ the radius) with large rotational velocity ($v$~$\sim$~225~km~s$^{-1}$)
can reproduce the data. In a subsequent paper however, Haehnelt et al.
\cite{haeh98} use hydrodynamic simulations in the framework of a standard
CDM cosmogony to demonstrate that the absorption profiles can be reproduced by
a mixture of rotational and random motions in merging protogalactic clumps.
The typical virial velocity of the haloes is about 100~km~s$^{-1}$. 

The issue of whether DLA systems originate in thick disks 
has been questioned previously. In particular,
the metallicity
distribution of DLA systems is inconsistent with that of stars in the thick
disk of our Galaxy (\cite{pett97b}; see however \cite{wol98}). 
Arguments in favor of DLA systems being associated with 
dwarf galaxies have been reviewed by Vladilo \cite{vlad98}. However,
it has been shown recently that DLA systems at intermediate redshift are
associated with galaxies of very different morphologies \cite{lebr97}. 
This strongly suggests that the
objects associated with high-redshift DLA absorbers are progenitors of
present-day galaxies {\sl of all kinds} (see \cite{led98}). 
%
\section{CONCLUSION}
The amount of information derived from studies of QSO absorption line systems
has increased tremendously during the last few years. This largely 
results from the improvements in instrumental sensitivity,
and spectral resolution and the extension of
the accessible wavelength range. Coupled with the development of 
sophisticated $N$-body simulations, this progress has 
favored the emergence of a consistent picture in which
not only the intergalactic medium is the baryonic reservoir for galaxy
formation but also galaxy formation strongly influences the evolution
of the IGM through metal enrichment and ionizing radiation 
emission.\par\noindent
One of the most exciting projects for the next few years may be to
probe the 3D spatial distribution of the gas at high redshift.
A number of large surveys
are underway or planed, which will provide a large number of bright
($m_{\rm V}$~$<$~19) quasars for investigation of large scales. 
A few small fields will be searched for deeper samples
of quasars ($m_{\rm V}$~$<$~23).
If a hundred lines of sight can be observed in one square degree,
the probability distribution function of the matter density could be 
investigated at high redshift. This would be a unique way to 
study how galaxy formation is related to the distribution and dynamics of the
underlying matter field.
%
%

%

%
%
%
\ack{\vskip -0.5cm
S. Charlot is thanked for a careful reading of the manuscript.}
%
%
%

%
%
%
%

\end{document}